\documentstyle[12pt]{article}
\def\beq{\begin{equation}}
\def\eeq{\end{equation}}

\def\beqn{\begin{eqnarray}}
\def\eeqn{\end{eqnarray}}
\relax
\begin{document}
\begin{titlepage}
\def\ba{\begin{array}}
\def\ea{\end{array}}
\def\thefootnote{\fnsymbol{footnote}}
\vfill
\hskip 4in BNL-60450, hep-ph9406324

\hskip 4in October, 1994
\begin{center}
{\large \bf BOUNDS ON $g_5^Z$ FROM PRECISION LEP
MEASUREMENTS}\\

\vspace{2 in}
{\bf S.~Dawson$^{(a)}$}{\bf  and G.~Valencia$^{(b)}$}\\
{\it  $^{(a)}$ Physics Department,
               Brookhaven National Laboratory,  Upton, NY 11973}\\
{\it  $^{(b)}$ Department of Physics,
               Iowa State University,
               Ames IA 50011}\\
\vspace{2 in}
\end{center}
\begin{abstract}

The parity violating but CP conserving anomalous three-gauge-boson
coupling $g_5^Z$ induces a universal contribution to the left-handed coupling
of the $Z$ boson to fermions. We find that the LEP measurements of
the partial $Z$ widths and lepton forward-backward asymmetries are
sufficiently precise to place a bound of order $|g_5^Z|$ less than
$\sim~10\%$.
This bound is significantly better than what can be obtained
at present from rare $K$ and $B$ meson decays.

\end{abstract}

\end{titlepage}

\clearpage

High precision measurements at the $Z$ pole at LEP combined with
polarized forward backward asymmetries at SLC put stringent
limits on any new physics beyond the standard
model.  The possible effects of new physics have typically
been parameterized in terms of the ``oblique'' corrections to
the standard model; those corrections involving the two-point functions
of the $W$ and $Z$ gauge bosons.  The ever increasing precision of the
electroweak measurements leads to the question of whether limits
can also be placed on the deviation of the three-point couplings
from their standard model values. The tree-level effects of these
couplings should be tested by future colliders. Here, we study their
effect on LEP observables at the one-loop level.

The effective Lagrangian formalism is particularly useful for this study.
In this approach, the low energy effects of potential new physics are
parameterized by a few coupling constants multiplying the lowest dimension
operators satisfying the symmetries of the standard model.

We have recently considered in some detail one of the next-to-leading
operators in the effective Lagrangian, the one responsible for the
``anomalous coupling'' $g_5^Z$ in the conventional parameterization
of the three-gauge-boson vertex. This is
the only operator that is parity violating, but CP
conserving. It contributes to the three and four gauge boson
vertices \cite{dv,cdhv}.
In this paper, we consider the bounds that can be placed on the
coupling constant $g_5^Z$ from precision measurements at LEP.

These bounds arise because at the one-loop level this operator
modifies the $Z f^\prime \overline{f}$ couplings. Because the operator
modifies the gauge boson self-couplings, its one-loop effects on
the $Z$ couplings to fermions affect both the flavor diagonal and the
flavor changing vertices. Since the
standard model does not contain flavor changing neutral vertices at
tree level, one can bound the size of $g_5^Z$ from rare meson decays.
Limits on the order $g_5^Z \leq {\cal O}(1)$ have been derived from
low energy decays such as $K_L\rightarrow \mu^+\mu^-$ \cite{dv,he}.
The flavor diagonal vertices have now been constrained by LEP to
such levels that the limits obtained from precision measurements
at LEP on $g_5^Z$ are more restrictive than those obtaining from
rare meson decays.

As has been discussed at length in the literature,
the minimal effective Lagrangian that describes the
interactions of gauge bosons of an $SU(2)_L \times U(1)_Y$ gauge
theory spontaneously broken to $U(1)_{EM}$ is given by:
\beq
{\cal L}^{(2)}={v^2 \over 4}D^\mu \Sigma^\dagger D_\mu \Sigma
-{1\over 2} {\rm Tr}\biggl(W^{\mu\nu}W_{\mu\nu}\biggr)
-{1\over 2}{\rm Tr}\biggl(B^{\mu\nu}B_{\mu\nu}\biggr)
+\Delta\rho{v^2\over 8} \biggl[Tr \biggl(
\tau_3 \Sigma^\dagger D_\mu \Sigma\biggr)\biggr]^2\quad ,
\label{lagt}
\eeq
where $W_{\mu\nu}$ and $B_{\mu\nu}$ are
the $SU(2)$ and $U(1)$  field strength tensors given in terms of
$W_\mu \equiv W^i_\mu \tau_i$,
\beqn
W_{\mu\nu}&=&{1 \over 2}\biggl(\partial_\mu W_\nu -
\partial_\nu W_\mu + {i \over 2}g[W_\mu, W_\nu]\biggr)
\nonumber \\
B_{\mu\nu}&=&{1\over 2}\biggl(\partial_\mu B_\nu-\partial_\nu B_\mu\biggr)
\tau_3.
\eeqn
The matrix $\Sigma \equiv \exp(i\vec{\omega}\cdot \vec{\tau} /v)$, contains the
would-be Goldstone bosons $\omega_i$ that give the $W$ and $Z$ their
masses via the Higgs mechanism and the $SU(2)_L \times U(1)_Y$
covariant derivative is given by:
\beq
D_\mu \Sigma = \partial_\mu \Sigma +{i \over 2}g W_\mu^i \tau^i\Sigma
-{i \over 2}g^\prime B_\mu \Sigma \tau_3.
\label{covd}
\eeq
The first term in Eq.~\ref{lagt} is the $SU(2)_L \times U(1)_Y$ gauge
invariant mass term for the $W$ and $Z$. The physical masses
are obtained with $v \approx 246$~GeV. This non-renormalizable Lagrangian
is interpreted as an effective field theory, valid below
some scale $\Lambda \leq 3$~TeV. The lowest order interactions between
the gauge bosons and fermions are the same as those in the minimal
standard model.

It is useful to classify the operators in the effective Lagrangian
into those that respect a ``custodial'' $SU(2)_C$ symmetry and those
that do not. In Eq.~\ref{lagt} the last term breaks the $SU(2)_C$
symmetry. The next-to-leading order effective Lagrangian has also been
discussed at length in the literature. It contains five terms
which conserve the custodial $SU(2)$ (up to hypercharge couplings)
and CP, six terms which
conserve CP, but break the custodial $SU(2)$, and three terms which
violate CP and the custodial $SU(2)$ \cite{long,appel}.
In this letter we will consider only one of these terms,
the one corresponding to a parity violating but $CP$ conserving operator.
This operator also breaks the custodial $SU(2)$ symmetry.
Because of the parity violating nature of this
term, it can lead to observable signals at LEPII via
$e^+e^-\rightarrow W^+W^-$\cite{dv} and at high energy $e\gamma$
colliders \cite{dv,cdhv}. The operator is:
\beq
{\cal L}^{(4)}=g \hat{\alpha} {v^2 \over \Lambda^2}
\epsilon^{\alpha \beta
\mu \nu}Tr \biggl(\tau_3 \Sigma^\dagger D_\mu \Sigma\biggr)
Tr\biggl(W_{\alpha \beta} D_\nu \Sigma \Sigma^\dagger\biggr)\quad ,
\label{lfour}
\eeq
where $\Lambda$ is the scale of the new physics responsible for this
operator, and our normalization is such that the coupling $\hat{\alpha}$
is naturally of order ${\cal O}(1)$ if the fundamental theory does not
have a custodial $SU(2)$ or of order ${\cal O}(\Delta\rho)$ if there is an
underlying custodial symmetry.
In terms of the conventionally used coupling:
\beq
g_5^Z = {g^2 \over c^2_\theta}\hat{\alpha} {v^2 \over \Lambda^2}
={4 M_Z^2 \over \Lambda^2} \hat{\alpha}
\quad .
\eeq
A term of this form is induced at the one-loop level
in the minimal standard model through fermion loops \cite{dv}. Beyond
the standard model it can occur, for example,
in technicolor models through techni-fermion loops \cite{appel}.
The operator of Eq. 4 modifies the minimal standard model $W^+W^-Z$
and $W^+W^-Z\gamma$ interactions. It does not modify the
$W^+W^-\gamma$ interaction.
We have presented the Feynman rules for this operator in unitary gauge
in Ref.~\cite{dv}.

The lowest order (standard model) coupling of a $Z$ to a fermion pair
can be written as
\beq
{\cal L}=-{ g\over 4 c_\theta} Z^\mu {\overline f}
\gamma_\mu \biggl(R_f(1+\gamma_5)+L_f (1-\gamma_5)\biggr)f
\eeq
where $R_f=-2 s^2_\theta Q_f$, $L_f=T_3-2s^2_\theta Q_f$, ($T_3=\pm 1$)
and $s^2_\theta\equiv
\sin^2\theta_W$.

When we abandon the renormalizable minimal standard model in favor of
a non-renormalizable effective field theory, we must in general allow
for additional gauge-boson fermion interactions. The next-to-leading terms
have been considered in Ref.~\cite{peccei}. One of the possible terms
listed in Ref.~\cite{peccei} is:
\beq
{\cal L}^{f}={g \over 2 c_\theta} \overline{f}_{Li} \gamma_\mu
(\kappa^{NC}_L)_{ij} f_{Lj} Z^\mu.
\label{pz}
\eeq
The notation is that of Ref.~\cite{peccei}. The index $L$ indicates that
this operator involves only the left-handed fermions (there are analogous
operators involving the right-handed fermions but we will not concern
ourselves with those). The $i,j$ are generation indices indicating that
in general this operator modifies the tree-level coupling $L_f$ and also
introduces flavor changing neutral couplings to the $Z$. The limits on
flavor changing neutral currents have been very stringent for a long time.
Those limits, as well as low energy measurements of $L_f$, taken from
Ref.~\cite{langacker} led the authors of Ref.~\cite{peccei} to conclude that
the $(\kappa^{NC}_L)_{ij}$ had to be less than 1\%.

Operators like Eq.~\ref{lfour} induce $(\kappa^{NC}_L)_{ij}$ at the
one-loop level, and thus at a natural size of a few percent or less. These
one-loop contributions to $(\kappa^{NC}_L)_{ij}$ are formally of next-
to-next-to-leading order. To reach precise conclusions from the study of
these effects one would need a complete effective field theory analysis
at next-to-next-to-leading order where the number of free parameters
becomes too large to be useful. However, we can bound three-gauge-boson
coupling constants like $g_5^Z$ from their contribution to
$(\kappa^{NC}_L)_{ij}$ under the naturalness assumption that no
cancellation will take place amongst the different contributions that can
occur at the same order.

We thus proceed to evaluate
the diagram of Fig. 1 to find the effect of $g_5^Z$ on the
$Z\overline{f}f$ coupling.  In order to preserve
gauge invariance we use dimensional regularization. Since, as
discussed above, we will not consider all the possible couplings
that occur at next-to-next-to-leading order (that would act as
counterterms for our loop calculations) we base our analysis on
the leading non-analytic contributions from the loop diagrams.
(This has become common practice in chiral perturbation theory
whenever a complete calculation with all possible counterterms is
not practical \cite{cpt}.)

We thus compute only the terms that go like $\log(\mu)$ for the
diagram shown in Fig.~1. These terms are easily found from the
coefficient of the divergent part of the integral (which is
dropped along with all other finite parts).
This gives us an estimate for the size of the new physics effects
if we choose the scale $\mu$ in such a way that
the $\log(\mu/M_W)$ is of order one.
We find  that this type of new physics affects only the left handed
coupling of the $Z$ to fermions and its effects can be incorporated
by modifying the tree level coupling in the form:
$L_f\rightarrow L_f + \eta c^2_\theta /s^2_\theta$ where
\beq
\eta={3\alpha \over 2\pi}g_5^Z \log\biggl({\mu\over M_W}\biggr).
\eeq
We have also neglected the mass and momentum of the external fermions
compared to the $Z$ mass. This modification of the lowest order
vertices, $\delta L_f$, does not grow with $M_t$ and is therefore
a universal contribution to all left-handed fermion interactions.

In addition to this direct contribution of $g_5^Z$ to the $Z f\overline{f}$
vertex we must consider indirect effects due to renormalization. In particular,
the operator of Eq.~\ref{lfour} also modifies the $W^\pm \rightarrow \ell^\pm
\nu$ coupling, contributing in this way to muon decay and thus introducing
a renormalization of $G_F$.\footnote{We are grateful to W.~Marciano for
bringing this point to our attention.} At the one loop-level (and always
working to lowest order in $g_5^Z$) we find from the diagrams in Figure~2
the following Lagrangian for
the effective $W^\pm \rightarrow \ell^\pm \nu$ coupling:
\beq
{\cal L}=-{ g\over 2\sqrt{2}} \biggl(1 - {\eta \over 2}\biggr)
W^\mu \overline{\ell}\gamma_\mu (1-\gamma_5 )\nu .
\eeq
We choose as our standard model input parameters: $G_F$ as measured in
muon decay, $\alpha_*(M_Z^2) \approx 1/128.8$ \cite{peskin}
and the physical $Z$ mass. We then
use a $s^2_\theta$ defined by the relation:
\beq
s^2_Z c^2_Z \equiv {\pi \alpha_* \over \sqrt{2} G_F M^2_Z} .
\eeq
Because of the structure of the $g_5^Z$ interaction (the epsilon tensor),
there are no one-loop contributions linear in $g_5^Z$ to any of the two
point functions and hence no wavefunction or $M_Z$ renormalization
linear in $g_5^Z$.  There is also no renormalization of $\alpha_*$
proportional to $g_5^Z$ since the $W^+W^-\gamma$ vertex is unaffected
by the interaction of Eq. 4.
 We are now in a position to compute the contribution of
the $g_5^Z$ term to the measured observables at LEP.

The extraordinary agreement between
the standard model predictions and the measurements at LEP allows us
to put very stringent bounds on the existence of the coupling
$g_5^Z$.  We begin by considering the partial decay widths.
For $Z \rightarrow f \overline{f}$ we write:
\beq
\Gamma (Z \rightarrow f \overline{f}) = \Gamma^{SM}_{f} + \delta\Gamma^5_{f}
\equiv
\Gamma^{SM}_{f}\biggl(1 + {\delta\Gamma^5_{f} \over
\Gamma^{(0)}_{f} } \biggr),
\eeq
where we have factored out the standard model result including radiative
corrections, $\Gamma^{SM}_f$. Since we drop higher order terms involving both
a standard model radiative correction and a correction due to $g_5^Z$, we
normalize the corrections introduced by $g_5^Z$, $\delta\Gamma^5_f$, to the
tree level partial width:
\beq
\Gamma^{(0)}(Z \rightarrow f \overline{f}) = N_c (R^2_f + L^2_f){G_F M_Z^3
\over 12 \pi \sqrt{2}},
\label{low}
\eeq
where $N_c=3$ ($1$) for quarks (leptons).
At the one-loop level the correction to the partial widths introduced by the
operator Eq.~\ref{lfour} is then:
\beq
{\delta\Gamma^{5}_{f} \over \Gamma^{(0)}_{f} }= \eta \biggl[
{2 L_f \over L_f^2 +R_f^2} {c^2_\theta \over s^2_\theta}
+ \biggl(1+{2 R_f(L_f + R_f) \over
L_f^2 +R_f^2}{c^2_\theta \over s^2_\theta - c^2_\theta}\biggr)\biggr]
\quad .
\label{modvertex}
\eeq
As we said, we find two contributions to $\delta\Gamma^5_{f}$.
The first term in Eq.~\ref{modvertex} is the direct
contribution from the induced $Z f \overline{f}$ vertex, and the
second term in Eq.~\ref{modvertex} comes from the
indirect contribution due to the renormalization of the tree-level
parameters in the lowest order result.

We compare the standard model predictions, $\Gamma_{f}^{SM}$,
including the one loop QED and QCD
radiative corrections with the most recent results from LEP.
We use the theory numbers of Langacker \cite{smpred} which use the global
best fit values for $M_t$ and $\alpha_s$ with $M_H$ in the range
$60-1000$~GeV. They include errors from the uncertainty in $M_Z$ and
$\Delta r$, $M_t$ and $M_H$, and from the uncertainty in $\alpha_s$.
 We add this theoretical uncertainty and the
experimental error in quadrature.
In all cases we will use $\mu =1$~TeV as the typical scale
in Eq. (8).
We present our results as $90\%$ confidence level intervals for the
allowed values of $g_5^Z$ in Table~\ref{t: pwid}. The sensitivity of
a particular observable to $g_5^Z$ depends on the combination of
couplings in Eq.~\ref{modvertex}. The most sensitive observable listed
in the table is $R_h$.

\begin{table}[htb]
\centering
\caption[]{$90\%$ confidence level intervals for $g_5^Z$ from different LEP
observables.}

\begin{tabular}{|c|c|c|c|} \hline
Observable & Experiment \cite{lepex} & SM prediction \cite{smpred}
 & $90\%$ c.l.
interval for $g_5^Z$ \\ \hline
$\Gamma_{ee}$ & $(83.98 \pm 0.18)$~MeV & $(83.87\pm 0.1)$~MeV
& (-0.10, 0.05) \\
$\Gamma_{\nu}$  & $(499.8\pm 3.5)$~MeV       & $(501.9\pm 0.91)$~MeV
& (-0.08,  0.04) \\
$R_h$         & $(20.795\pm0.040)$  & $20.782\pm 0.031$ & (-0.07, 0.10) \\
$\Gamma_{Z}$ & $(2497.4 \pm 3.8)$~MeV & $(2496 \pm 4.4)$~MeV & (-0.9, 1.2)
\\ \hline
\end{tabular}
\label{t: pwid}
\end{table}

The limits presented in Table~\ref{t: pwid}~ are more than an order of
magnitude better than the limit
$g_5^Z < {\cal O}(1)$ obtained from rare $K$ and $B$ decays \cite{dv}.
They are comparable to the limits that could be placed
in future high energy $e^- \gamma$ colliders \cite{cdhv}.

We have not included in Table~\ref{t: pwid} the $Z\rightarrow b \overline{b}$
width. Presented as a ratio to the total hadronic width of the $Z$,
the latest experimental result is $R_{bh}=0.2202 \pm 0.0020$ \cite{lepex}.
The theoretical prediction is $\delta_{bb}^{new}=.022\pm .011$
\cite{smpred}, where $\delta_{bb}^{new}=[\Gamma(Z\rightarrow b{\overline b})
-\Gamma(Z\rightarrow b {\overline b}^{(SM)})]
/\Gamma(Z\rightarrow b {\overline b}^{(SM)})$,
which falls outside the
experimental $90\%$ c.l. range. A negative value of $g_5^Z$ would push
the theoretical prediction up and we can use this to obtain
$|g_5^Z|\leq 0.7$ at $90\%$ c.l. which is not competitive with the
limits presented in Table~\ref{t: pwid}.

We can also extract a limit on $g_5^Z$ from forward backward asymmetries
at LEP. The new interaction proportional to $g_5^Z$ modifies the
asymmetries in a way most easily parameterized by:
\beq
{g_V\over g_A}=\biggl({g_V\over g_A}\biggr)_{SM}\biggl[
1+\biggl({2 R_\ell \over R_\ell^2-L_\ell^2}{c^2_\theta \over s^2_\theta}
+{2 R_\ell \over R_\ell +L_\ell}{c^2_\theta \over s^2_\theta - c^2_\theta}
\biggr)\eta \biggr]
\label{asym}
\eeq
where again the first correction term comes from the direct $Z f \overline{f}$
vertex and the second one from the renormalization of the tree level
parameters.
With the current theoretical results for the effective $g_V / g_A$, and
experimental value, the bounds placed on $g_5^Z$ are not as good as those
coming from the partial $Z$ widths. Using the numbers in
Ref.~\cite{ltasi} for example, we find the $90\%$ c. l. interval
for $g_5^Z$: $-0.3 \leq g_5^Z \leq 0.1$.

In conclusion we have found that precision measurements of the partial
$Z$ widths at LEP constrain the anomalous three-gauge-boson coupling
$g_5^Z$ to be of order $|g_5^Z|$ less than $\sim~10\%$. This is
significantly
better than the bounds that can be placed on $g_5^Z$ from rare $K$ and $B$
meson decays. It is comparable to the bounds that one will be able to place
on this coupling at future $e \gamma$ colliders.
These limits, however,
should be taken only as approximate indications of the
sensitivity of the LEP experiments to the presence of the $g_5^Z$ coupling.
More precise statements would require a complete effective field theory
calculation at next-to-next-to leading order.

\noindent{\bf Acknowledgments}

\noindent The work of S. Dawson is
supported by the U.S. DOE under contract DE-AC02-76CH00016.
We are grateful for conversations with W.~Marciano and A.~Sirlin.
\appendix

\noindent{\bf FIGURE CAPTIONS}

\begin{enumerate}

\item  One-loop diagram giving a contribution from the
anomalous three-gauge-boson coupling $g_5^Z$
to the $Z {\overline f} f$ vertex.

\item One-loop diagrams giving a contribution from the
anomalous three-gauge-boson coupling $g_5^Z$ to the
$W^\pm \ell^\pm \nu$ vertex.

\end{enumerate}

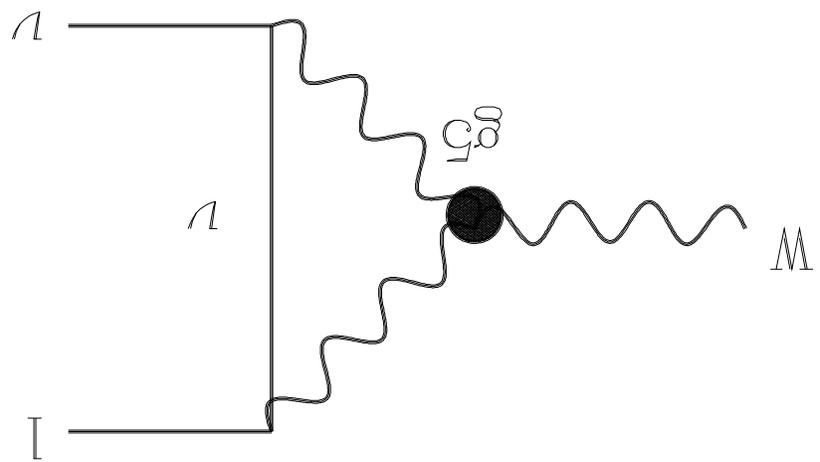
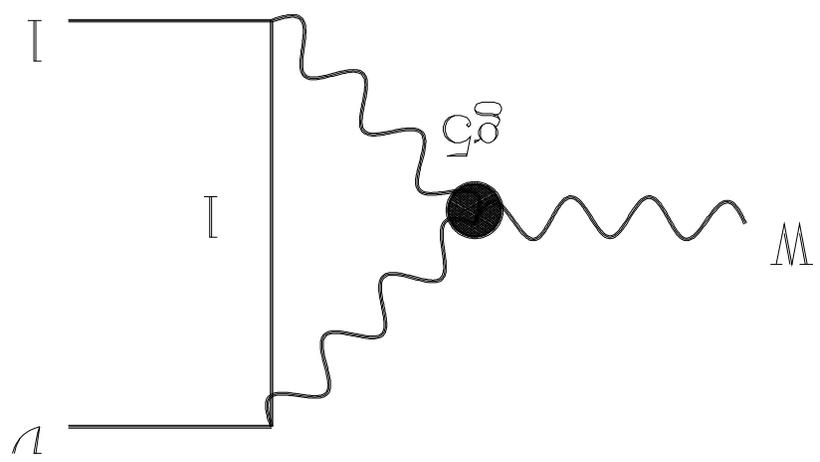
\end{document}